\documentclass[aps,physrev,preprint,groupedaddress]{revtex4-2}

\usepackage{graphicx}
\usepackage{dcolumn}
\usepackage{bm}

\usepackage[utf8]{inputenc}
\usepackage[T1]{fontenc}
\usepackage{mathptmx}
\usepackage{etoolbox}

\usepackage{siunitx}
\usepackage{lipsum}
\DeclareSIUnit\torr{Torr}
\DeclareSIUnit\oersted{Oe}
\usepackage{amsmath}

\begin{document}


\title{Thickness-Dependent Spin Pumping in YIG/$W_{90}Ti_{10}$ Bilayers}


\author{M. Hachem}
\author{Z. Harajli}
\author{S. Isber}
\author{M. Haidar}
\email{mh280@aub.edu.lb}
\affiliation{Department of Physics, American University of Beirut, Riad El-Solh,
Beirut, 1107-2020, Lebanon }%



\date{\today}

\begin{abstract}
We investigate the spin pumping efficiency in $YIG/W_{90}Ti_{10}$ bilayers by measuring the thickness dependence of both the YIG and WTi layers using broadband ferromagnetic resonance (FMR) spectroscopy. The deposition of a 5-nm WTi layer leads to enhanced Gilbert damping in thinner YIG films, indicating efficient spin current injection. From the spin pumping contribution to the damping of the YIG/WTi bilayer, we determine an effective spin mixing conductance of $3.3 \times 10^{18}~\mathrm{m^{-2}}$ for the 5-nm WTi layer. Further measurements with varying WTi thickness reveal a non-monotonic dependence of spin mixing conductance, peaking at $4.2 \times 10^{18}~\mathrm{m^{-2}}$ for a 3-nm WTi layer. This behavior is attributed to a structural phase transition from the high-spin–orbit $\beta$-phase to the less efficient $\alpha$-phase in thicker WTi layers. Furthermore, comparative analysis with YIG/W bilayers shows that Ti doping significantly reduces $g^{\uparrow \downarrow}_{\text{eff}}$. These findings highlight the critical role of alloy composition and structural phase in tuning spin transport for spintronic applications.

\end{abstract}


\maketitle

\section{Introduction}

The generation and injection of pure spin currents—i.e., spin angular momentum transport without an accompanying charge current— generated via the spin Hall effect \cite{Hirsh1999,dyakonov1971} has become increasingly critical for potential applications of spintronic devices \cite{Dieny2020}.  Bilayer structures comprising a ferromagnetic (FM) layer and a nonmagnetic (NM) layer with strong spin–orbit coupling represent promising platforms for the conception of tunable spin currents via the spin pumping mechanism \cite{Tserkovnyak2005,Kajiwara2010,Ando2011,Balinsky2015}.  In spin pumping, spin currents are dynamically injected from the FM layer into the NM layer through the excitation of magnetization precession, achieved via ferromagnetic resonance (FMR) \cite{Mosendz2010,Shaw2012}. Recently, extensive research works have investigated the spin pumping behavior in FM/NM heterostructures employing both metallic \cite{Nan2015,Ranjbar2014,Liu2011,Haidar2021} and insulating ferromagnets\cite{Wang2013,Sun2013,Heinrich2011,Haidar2016,Haidar2015}. 
In these devices, it is essential to determine key parameters include the spin Hall angle ($\theta_{SH}$) \cite{Wang,RojasSanchez2014,Rezende2013}, which quantifies the efficiency of spin-to-charge current conversion, and the inverse spin Hall voltage \cite{Yoshino2011,Liu2012,Balinsky2016,Dhananjay2017,Sagasta2018}, which reflects the magnitude of charge current generated from the injected spin current via spin–orbit mechanisms. The transport of the spin current is also known to be strongly influenced by the structural and interfacial properties of the FM/NM interface\cite{Nakayama2012, Jungfleisch2,HARAJLI2024}. The spin mixing conductance ($g^{\uparrow\downarrow}_\text{eff}$), which characterizes the efficiency of spin angular momentum transfer across the FM/NM interface, is another important parameter in the current conversion \cite{Tserkovnyak2002,Zhang2011}.
Several studies conducted to enhance the spin mixing conductance $g^{\uparrow\downarrow}_\text{eff}$ to increase the spin current density transmitted into nonmagnetic (NM) materials \cite{Papaioannou2013,Wang2019}. Notable improvements have been achieved by modifying the growth conditions such as temperature \cite{Bansal2018}, oxygen pressure \cite{Demasius2016}, introducing an ultrathin interlayer(e.g. Copper or hafnium \cite{Pai2014,Mazraati2018}or anti-ferromagnet \cite{WangAFM,Hahn2014}), engineering the interface morphology \cite{Liu2011,Jungfleisch2}. 
Beyond traditional heavy metals, research has also investigated materials with varying spin–orbit coupling strengths, such as alloying heavy metals $NixCu_{1-x}$\cite{Kelller2019}, PtAu \cite{Obstbaum2016,Zhu2018} to strengthen the generation of spin current. Tungsten is one of the most widely used heavy metals in spintronic devices due to its superior properties compared to other heavy metals. It has been shown that the spin Hall angle is significantly enhanced when tungsten is in the $\beta$-phase \cite{Mazraati2016,Zahedinejad2018apl,Bansal2018,Jhajhria2019}, and when it is oxidized \cite{Demasius2016}. The quest to enhance spin pumping in tungsten remains ongoing, and one approach has involved alloying W with other metals such as Tantalum (Ta), Copper, or Molybdenum \cite{Sui2017,Kim2020}.

Here, we explore spin pumping in titanium-doped tungsten ($W_{90}Ti_{10}$)  as a heavy metal layer. Using broadband ferromagnetic resonance (FMR), we systematically investigate YIG/WTi bilayers by varying both the YIG and WTi layer thicknesses. Our measurements reveal that spin pumping is enhanced in thinner YIG films. Furthermore, the spin-mixing conductance exhibits a non-monotonic dependence on WTi thickness, reaching a maximum value of $g^{\uparrow\downarrow}_\text{eff} = 4.2 \times 10^{+18} m^{-2}$ at a WTi thickness of 3 nm which is attributed to a phase transition from the high-spin–orbit $\beta$ -phase to the less efficient $\alpha$-phase of WTi for thicker layers.

\begin{figure*}[!h]
\begin{center}
\includegraphics[width=0.65\textwidth]{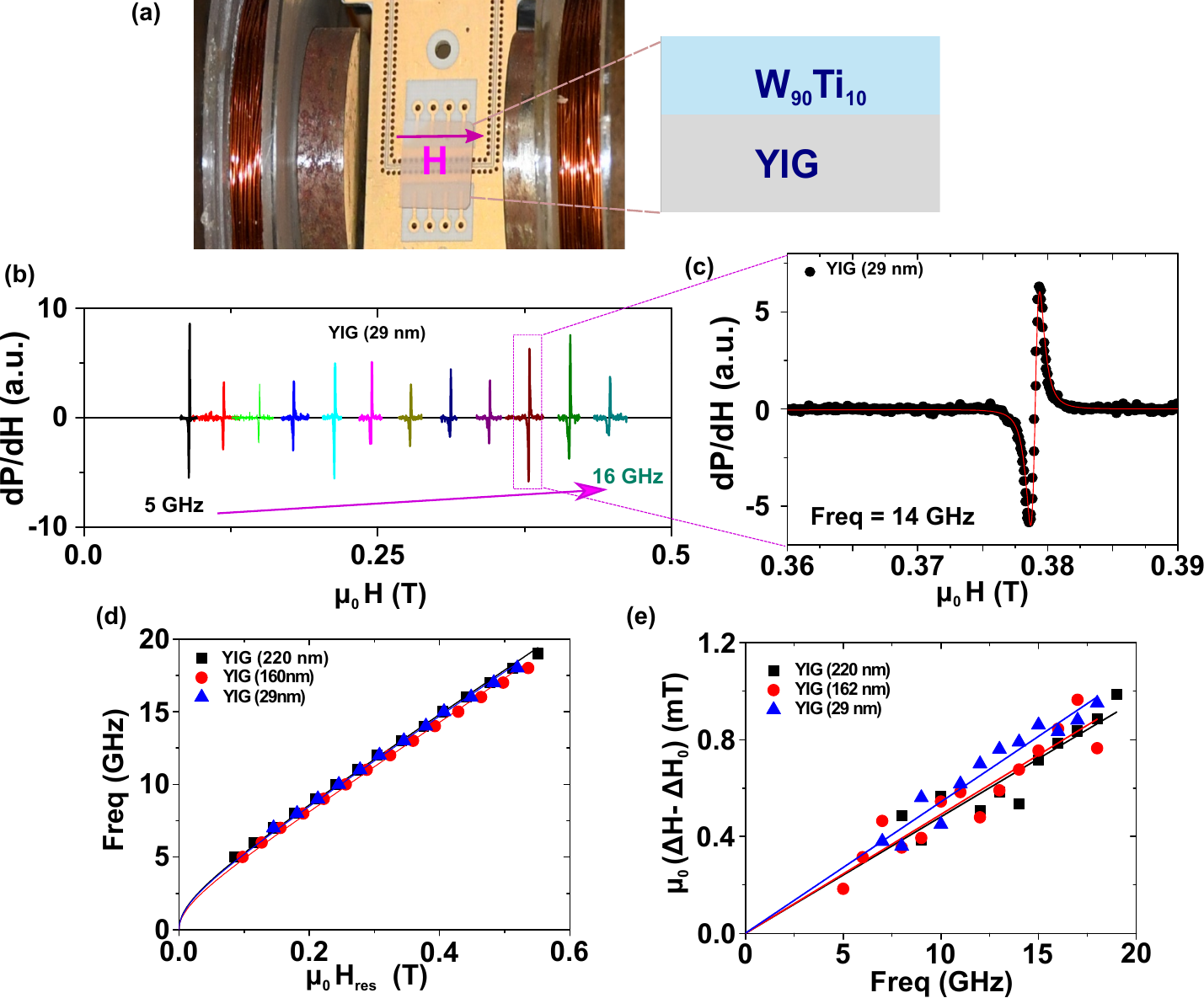}
\caption{(Color online) 
(a) (a) Schematic illustration of the broadband ferromagnetic resonance (FMR) measurement setup, showing the thin-film sample orientation, the applied in-plane magnetic field ($\mu_0 H$), and the YIG/WTi bilayers.
(b) Representative FMR spectra of a YIG film recorded at frequencies ranging from 5 to \SI{16}{GHz} in \SI{1}{GHz} increments. (c) Zoomed-in FMR spectrum at \SI{14}{GHz} with a fitted curve (solid line) using Eq. (1). (d) Frequency-field dispersion relation for all samples.(e) Variation of linewidth versus frequency for all pure YIG samples.}
\end{center}
\end{figure*}

\section{Experiments}
We selected yttrium iron garnet (YIG , $Y_3Fe_5O_{12}$) as the ferromagnetic (FM) layer due to its ultra-low magnetic damping. YIG thin films were grown on single-crystal (111)-oriented gadolinium gallium garnet (GGG, $Gd_3Ga_5O_{12}$) substrates using pulsed laser deposition (PLD). A YIG target was ablated with a 248 nm KrF excimer laser operating at 400 mJ pulse energy and 10 Hz repetition rate. Depositions were performed at a substrate temperature of 750 $^\circ$C in 20 mTorr oxygen pressure, with a base chamber pressure of $10^{-7}$ mbar. More details about the deposition protocol can be found \cite{Haidar2023}. 
Two series of YIG films were prepared. In Series I, the YIG film thicknesses were varied by adjusting the deposition time to achieve different thicknesses ranging from 20 to 220 nm. The thickness of the YIG films was determined using a profilometer. In Series II, all films were grown with identical thicknesses by fixing the deposition time at 50 minutes. For the spin pumping study, a thin ($W_{90}Ti_{10}$) alloy layer of variable thickness was sputtered onto each YIG film. We examine the effect of Ti doping in the W matrix to better understand its influence on the spin pumping of the bilayer. 

We conducted broadband ferromagnetic resonance (B-FMR) measurements on $5 \times 5$ $\mathrm{mm}^2$ rectangular samples using a coplanar waveguide to generate microwave excitations. We mounted the samples in a flip-chip configuration and applied the external magnetic field in-plane, as illustrated in Fig. 1(a). Using a lock-in amplifier, we detected the FMR signals while sweeping the magnetic field at fixed frequencies. Measurements were performed from \SI{3}{}--\SI{16}{GHz} at \SI{0}{dB} input power. The spectra, recorded as the derivative $(\mathrm{d}P/\mathrm{d}H)$, exhibit the characteristic FMR lineshape shown in Fig. 1(b,c). We fitted the FMR signal to the derivative of a Lorentzian function that includes both symmetric and antisymmetric components,

\begin{equation}
\begin{split}
  \frac{dP}{dH} &=A\frac{-8 \Delta H^2 \cdot (H - H_{\text{res}})}{(4(H - H_{\text{res}})^2 + \Delta H^2)^2} \\
  &+ B\frac{ \Delta H \cdot (\Delta H^2 - 4(H - H_{\text{res}})^2)}{(4(H - H_{\text{res}})^2 + \Delta H^2)^2} 
   + C \cdot H + D
\end{split}
\end{equation}
where $H_{\mathrm{res}}$ and $\Delta H$ are the resonance field and the linewidth, $A$ and $B$ represent the amplitude of symmetric and anti-symmetric terms, and $C$ and $D$ are the slope and constant factor, respectively.

\begin{table*}[!hb]
\centering
\begin{tabular}{c|c|c|c|c}
\hline
\textbf{Sample} & YIG thickness & $\mu_0 M_s$ & $\alpha_{YIG}$ & $\alpha_{YIG/WTi}$  \\
& (nm) & (T) & ($\times 10^{-4}$) & ($\times 10^{-3}$) \\
\hline
YIG / WTi (5 nm) & 29 & 0.2 & 8.1 & 2  \\
YIG / WTi (5 nm) & 162 & 0.16 & 7.9 & 1.2 \\
YIG / WTi (5 nm) & 220 & 0.21 & 7.2 & 1.07  \\
YIG / WTi (1 nm) & 26 & 0.2 & 10.2 & 1.4 \\
YIG / WTi (3 nm) & 27 & 0.2 & 6.5 & 2.2  \\
YIG / WTi (7 nm) & 29 & 0.19 & 8.6 & 1.45  \\
YIG / WTi (10 nm) & 26 & 0.2 & 8.5 & 1.65 \\
\hline
\end{tabular}
\caption{Extracted values of the saturation magnetization ($\mu_{0}M_{\mathrm{s}}$), intrinsic damping constant ($\alpha_{\mathrm{YIG}}$) for YIG films, and total damping ($\alpha_{\mathrm{YIG/WTi}}$) for YIG/WTi bilayers.}
\label{tab:3.3}
\end{table*}

\section{Results}
We analyze the resonance field and linewidth to assess the magnetodynamic properties of the YIG films. The saturation magnetization ($\mu_{0}M_{\mathrm{s}}$) was determined by fitting the frequency–field dispersion, shown in Fig. 1(d), to Kittel’s equation 
\begin{equation}
\textit{f}=\mu_{0}\gamma/2\pi \sqrt{(H)(H+M_{s})}, 
\end{equation}
where $\gamma/2\pi$ is the gyromagnetic ratio of \SI{30}{GHz/T}. Analyzing the frequency dependence of the FMR linewidth ($\mu_{0}\Delta H$) enabled the extraction of the Gilbert damping constant ($\alpha$), as shown in Fig. 1(e). Across all samples, the linewidth exhibited a linear increase with frequency from 2 to 16 GHz, reflecting its direct proportionality to energy dissipation. The total FMR linewidth includes both intrinsic and extrinsic contributions and is described by:
\begin{equation}
\mu_0\Delta H_{} = \mu_0 \Delta H_0 + (2\alpha/\gamma)2\pi f,
\end{equation}
where $\mu_0 \Delta H_0$ is the inhomogeneous broadening. 

Table 1 summarizes the magnetodynamic properties of the YIG films, including $\mu_{0}M_{\mathrm{s}}$ and $\alpha$.

We study the spin pumping effect in YIG/WTi as a function of the YIG layer thickness. A 5-nm WTi layer is sputtered onto YIG films with thicknesses of 29, 162, and 220 nm. An example of the FMR signal measured for YIG (29 nm)/WTi (5 nm) bilayers is shown in Fig. 2(a). One can observe a broad peak for the bilayers, indicating an enhanced damping compared to the YIG films alone, mainly due to the spin pumping effect. In bilayers, the total damping $\alpha_{YIG/WTi}$ is the sum of the intrinsic damping of the YIG film ($\alpha_{YIG}$) and the spin pumping contributions to the damping ($\alpha_{sp}$), 
\begin{equation}
   \alpha_{YIG/WTi} = \alpha_{YIG} + \alpha_{sp}. 
   \label{eq:sp}
\end{equation} 
Fig. 2(b) shows the variation of the linewidth versus frequency for YIG/WTi bilayers with different YIG layer thicknesses. One can notice that the slope of $\Delta H/f$ decreases with the increase of YIG thickness. The inset of Fig. 3(a) shows the variation of $\alpha_{YIG}$ and $\alpha_{YIG/WTi}$ as a function of YIG thickness, where the damping decreases for thicker films. 
Fig. 3 shows the variation of $\alpha_{sp}$, calculated from Eq. \ref{eq:sp}, as a function of $t_{YIG}$, where an increase in spin pumping is observed for thinner YIG layers, reaching a maximum of $1.25 \times 10^{-3}$ for the 29 nm thick YIG films. This is a typical result observed in ferromagnet/ heavy metals \cite{Heinrich2011, Jungfleisch2013,Haertinger2015}. Due to the conservation of angular momentum, the $\alpha_{sp}$ provides a means to evaluate the effective spin mixing conductance ($g^{\uparrow \downarrow}_{\text{eff}}$) at the YIG/WTi interface which quantifies the efficiency of spin angular momentum transfer across the interface between a ferromagnetic and the heavy metals, by
\begin{equation}
\alpha_{sp} = \frac{g \mu_{B} }{4\pi M_{s}t_{YIG }} g^{\uparrow \downarrow}_\text{eff}
\label{asp}
\end{equation}

\begin{figure*}[!h]
\begin{center}
\includegraphics[width=0.7\textwidth]{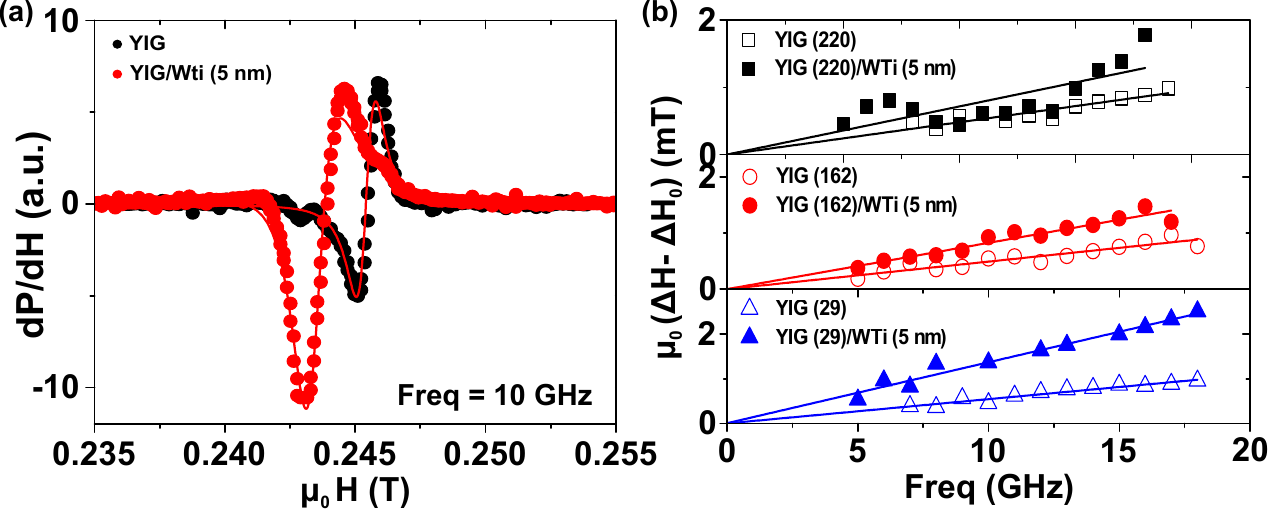}
\caption{(Color online) (a) FMR spectra of a YIG film measured at \SI{10}{GHz}, shown for the film without a WTi layer (black symbols) and with a WTi layer (red symbols). The solid red line represents the fit to Equation (1).
(b) Variation of linewidth versus frequency for YIG (t)/WTi (5 nm) samples with different YIG thicknesses. Solid lines represent fits to Eq. (3).}.
\end{center}
\end{figure*}

\begin{figure}[!h]
\begin{center}
\includegraphics[width=0.40\textwidth]{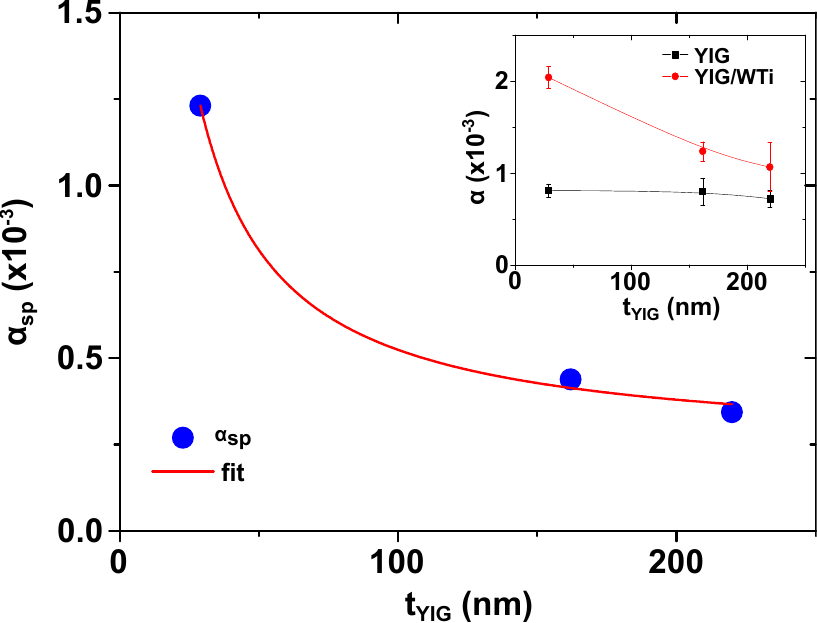}
\caption{(Color online) Variation of the spin pumping damping ($\alpha_{sp} = \alpha_{YIG/ WTi} - \alpha_{YIG}$) as a function of YIG thickness. Insets: Variation of $\alpha_{yig}$ and $\alpha_{yig/ WTi}$ with YIG thickness. Solid lines serve as guides to the eye. }
\end{center}
\end{figure}

where $g$, $\mu_B$, $M_s$, and $t_{\text{YIG }}$ represent the Landé g-factor, Bohr magneton, saturation magnetization, and thickness of the YIG layer, respectively. Here, $g^{\uparrow \downarrow}_{\text{eff}}$ accounts for the spin current backflow driven by spin accumulation. By fitting the data in Fig. 3, we extract an effective spin mixing conductance $g^{\uparrow \downarrow}_{\text{eff}}$ of $3.3 \times 10^{18} m^{-2}$ for the 5 nm WTi layer.

To investigate the influence of WTi layer thickness on spin pumping, we prepared a series of YIG films with comparable thicknesses and similar magnetic properties, as summarized in Table 1. A WTi layer with varying thicknesses ranging from 1 to 10 nm was subsequently sputtered onto each YIG film. Fig. 4(a) shows the frequency dependence of the FMR linewidth for bare YIG films (black squares) and YIG/WTi bilayers (colored symbols). A steeper slope is observed for the YIG/WTi (3 nm) bilayer, indicating enhanced damping. The spin pumping contribution, calculated as discussed above using Eq. \ref{eq:sp}, is plotted in the inset of Fig. 4(b). 
A non-monotonic dependence of spin pumping on WTi thickness is observed, with a maximum at 3 nm followed by a decrease at higher thicknesses. The effective spin-mixing conductance, $g^{\uparrow \downarrow}_{\text{eff}}$, was then calculated from the spin pumping contribution. As shown in Fig. 4(b), $g^{\uparrow \downarrow}_{\text{eff}}$ increases from $0.5 \times 10^{18} m^{-2}$ at 1 nm to a maximum of $4.2 \times 10^{18} m^{-2}$ at 3 nm. This behavior is commonly observed, where spin pumping increases for heavy metal thicknesses below the spin diffusion length \cite{Sun2013}. At higher thicknesses, the spin-mixing conductance decreases, reaching $1\times 10^{18} m^{-2}$ at 10 nm. A comparison between WTi (3 nm) and WTi (10 nm) reveals a four times increase in $g^{\uparrow \downarrow}_{\text{eff}}$ with decreasing thickness, suggesting that spin current generation is most efficient in the YIG/WTi (3 nm) bilayer. This can be attributed to the stabilization of the $\beta$-phase in thinner WTi layers, which is associated with strong spin–orbit coupling. In contrast, thicker films tend to favor the  $\alpha$-phase, leading to a reduced spin pumping efficiency. Four-point probe resistivity measurements indicate that the films are highly resistive, suggesting partial oxidation of the WTi layer. This is further supported by energy-dispersive X-ray (EDX) analysis, which reveals an oxygen content of approximately 15\% in the WTi layers. Notably, this observation aligns with findings from a recent study, which demonstrated that oxygen is shown to stabilize the $\beta$-W structure by occupying interstitial sites and altering the local bonding, where a clear correlation between the presence of oxygen and the formation of the $\beta$-phase is established both experimentally and theoretically \cite{Biswas2020}. These results suggest that oxygen incorporation in our WTi films may also play a role in stabilizing the $\beta$-phase at lower thicknesses.

\begin{figure*}[!h]
\begin{center}
\includegraphics[width=0.8\textwidth]{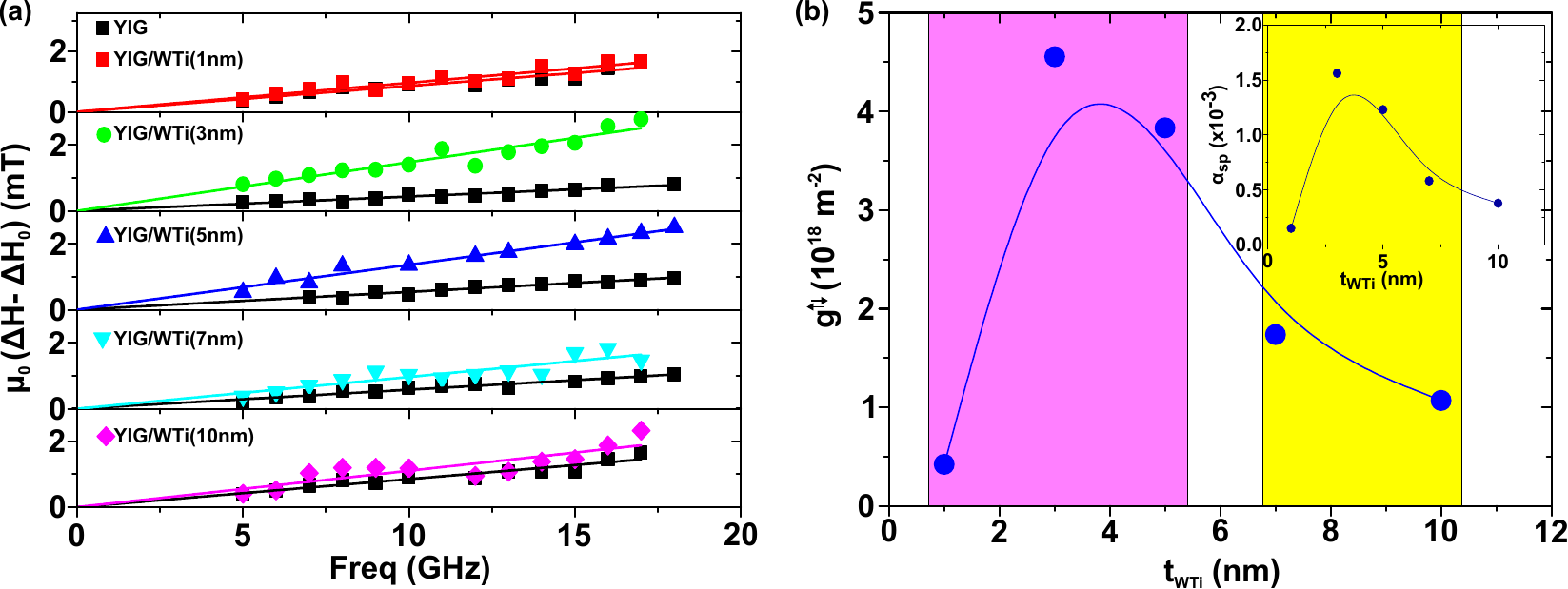}
\caption{(Color online) (a) Variation of the FMR linewidth as a function of frequency for YIG films (black squares) and YIG (30 nm)/WTi bilayers (colored symbols) with varying WTi thicknesses. (b) Effective spin-mixing conductance $g^{\uparrow \downarrow}_{\text{eff}}$ as a function of WTi thickness. The inset shows the corresponding spin-pumping contribution to the damping $\alpha_{sp}$ versus WTi thickness.}
\end{center}
\end{figure*}

To study the effect of Ti doping in the W matrix and its influence on spin pumping, we compare the spin transport characteristics of YIG (30 nm)/WTi (10 nm) and YIG (30 nm)/W (10 nm) bilayers. For this purpose, we deposit a 10 nm-thick W layer on top of the YIG film. Fig. 5 shows the variation of the FMR linewidth as a function of frequency for YIG films with and without W (top panel) and WTi (lower panel) layers. As discussed above, we extract the spin pumping contribution to the damping, $\alpha_{sp}$, of $9 \times 10^{-4}$ for the YIG/W bilayer and of $3.8 \times 10^{-4}$ for the YIG/WTi. We estimate the effective spin mixing conductance, $g^{\uparrow \downarrow}_{\text{eff}}$, of $2.93 \times 10^{18}\mathrm{m^{-2}}$ for the YIG/W bilayer and of $1.07 \times 10^{18}\mathrm{m^{-2}}$ for the YIG/WTi bilayers. Our results highlight the strong influence of alloying on interfacial spin transport, as evidenced by about 50\% decrease in the spin-mixing conductance upon introducing Ti into the W matrix, compared to YIG/W (10 nm) films. This behavior may be attributed to the low atomic number of Ti, which exhibits relatively weak spin–orbit coupling, thereby reducing the overall spin–orbit interaction in the WTi alloy and diminishing the efficiency of spin current. Importantly, such alloying offers a versatile route to engineer spintronic interfaces. For instance, first-principles calculations by Sui \textit{et al.} reveal that substituting Tantalum (Ta) into $\beta$-W markedly enhances the spin Hall conductivity for $W_{0.875}Ta_{0.125}$, suggesting improved spin transparency and elevated spin-mixing conductance in these systems \cite{Sui2017}. Similarly, in $W_{x}Mo_{1-x}$ / YIG bilayers, tailoring the W-Mo ratio enables stabilization of favorable crystalline phases, thereby optimizing spin pumping efficiency for spintronic applications \cite{Ma2018}. Furthermore, alloying tungsten (W) with copper (Cu) has been shown to effectively tune the spin Hall angle, with a peak observed at a W concentration of 60\%, exceeding that of pure W \cite{COESTER2021}.

\begin{figure}[!h]
\begin{center}
\includegraphics[width=0.4\textwidth]{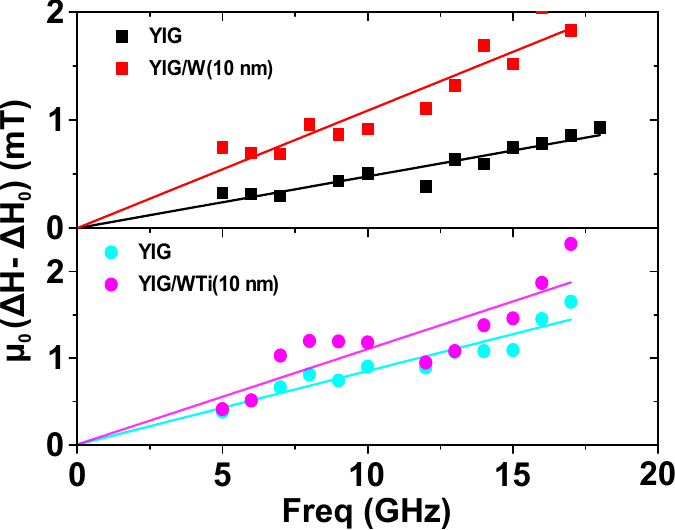}
\caption{(Color online) Variation of linewidth versus frequency for YIG (30 nm)/WTi (10 nm) in the lower panel and YIG (30 nm)/W (10 nm) in the upper panel.}
\end{center}
\end{figure}

\section{Conclusion}
In conclusion, we investigated the thickness dependence of spin pumping in YIG/WTi bilayers by employing broadband ferromagnetic resonance (FMR) spectroscopy. Analysis of the damping behavior revealed enhanced spin pumping in thinner YIG layers, with an effective spin mixing conductance of $3.3 \times 10^{18}~\mathrm{m^{-2}}$ for a 5-nm-thick WTi layer. Varying the WTi thickness produced a non-monotonic trend in spin pumping efficiency, peaking at $4.2 \times 10^{18}~\mathrm{m^{-2}}$ for a 3-nm WTi layer. This behavior is attributed to a phase transition from the high-spin–orbit $\beta$-phase to the less efficient $\alpha$-phase at larger thicknesses. Additionally, comparison with YIG/W bilayers revealed that Ti incorporation reduces the spin mixing conductance by over 50\%, consistent with the weaker spin–orbit coupling of Ti. These findings highlight the critical role of alloy composition, crystalline phase, and thickness control in modulating interfacial spin transport for spintronic device engineering.

\textbf{ACKNOWLEDGMENT}
\\
The authors would like to acknowledge the financial support of the American University of Beirut Research Board (URB).

\textbf{DATA AVAILABILITY}

The data that support the findings of this study are available from the corresponding author upon reasonable request.

%



\end{document}